\documentclass{article}
\usepackage{spconf,amsmath,amsfonts,graphicx,multirow}
\usepackage[caption=false]{subfig}

\title{PLUMBERNET: FIXING INTERFERENCE LEAKAGE AFTER GEV BEAMFORMING}
\name{Fran\c{c}ois Grondin\textsuperscript{1}, Caleb Rasc\'on\textsuperscript{2}}
\address{\textsuperscript{1}Universit\'e de Sherbrooke, Sherbrooke (Qu\'ebec), Canada,\\\textsuperscript{2}Universidad Nacional Aut\'onoma de M\'exico, Ciudad de M\'exico, M\'exico}
\begin{document}
%
\maketitle
\begin{abstract}
Spatial filters can exploit deep-learning-based speech enhancement models to increase their reliability in scenarios with multiple speech sources scenarios. To further improve speech quality, it is common to perform postfiltering on the estimated target speech obtained with spatial filtering. In this work, Generalized Eigenvalue (GEV) beamforming is employed to provide the leakage estimation, along with the estimation of the target speech, to be later used for postfiltering. This improves the enhancement performance over a postfilter that uses the target speech and a reference microphone signal. This work also demonstrates that the spatial covariance matrices (SCMs) can be accurately estimated from the direction of arrival (DoA) of the target and a discriminative selection amongst the pairwise estimated time-frequency masks.
\end{abstract}
\begin{keywords}
Beamforming, speech, postfilter, leakage
\end{keywords}
\section{Introduction}
\label{sec:intro}

In real-world case scenarios, speech is captured in conjunction with background noise and other interferences.
Speech enhancement aims to eliminate both noise and interferences from the captured signal to only keep the target speech. This field has gained significant attention lately, largely due to the impressive advancements in deep learning models \cite{das2021fundamentals}.
Most approaches rely on single-channel input \cite{lu2023mp,kong2023cleanunet}, though multi-microphone solutions can improve performances as they also exploit the spatial information.
While end-to-end multichannel models can achieve high quality speech enhancement, they rely on multi-microphone device-specific datasets, which makes data collection challenging \cite{kraaij2005ami}.
For these reasons, a two-stage strategy that involves multi-channel spatial beamforming followed by single channel speech enhancement makes the processing agnostic to the microphone array shape \cite{wang2021sequential,rascon2023target}.
Beamforming methods often exploit deep neural networks that generate time-frequency mask to estimate accurate spatial covariance matrices (SCMs) \cite{erdogan2016improved,heymann2015blstm}.

Recently, a Guided Speech Enhancement Network was introduced as a dual-input speech enhancement neural network that uses the Minimum Variance Distortionless Response (MVDR) beamformer output and a reference microphone \cite{yang2023guided}.
Results demonstrate the effectiveness of the method when the target speech is competing with interfering speech.
However, this method requires \emph{a priori} knowledge of the MVDR coefficients, that are estimated from short utterances played at each desired direction of interest, which makes the approach less suitable for dynamic environments.

This paper presents two contributions: 1) we propose a microphone array geometry agnostic method to estimate the SCMs using the target direction of arrival and the former SteerNet model \cite{grondin2020gev}; and 2) inspired by classical microphone array post-filtering strategies that use leakage estimate \cite{valin2004microphone,maldonado2020lightweight}, we demonstrate that using the estimated interference signal (i.e. `leakage') rather than a reference microphone as the second input to the neural network improves target speech enhancement performances when using the same neural network architecture and size. 

\section{Proposed Method}
\label{sec:method}

\begin{figure*}
    \centering
    \includegraphics[width=\linewidth]{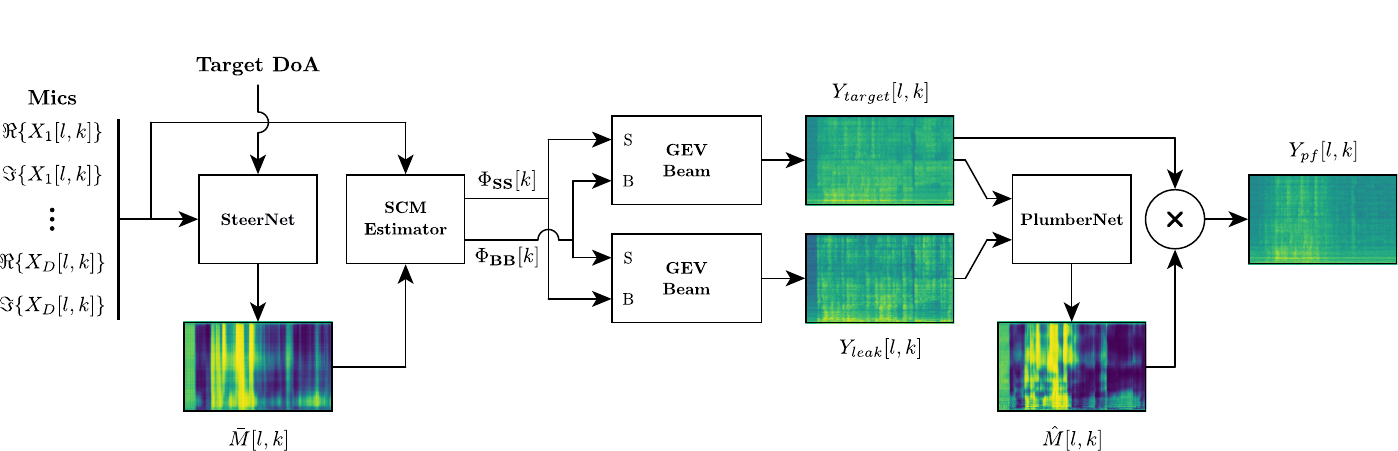}
    \caption{Proposed method. SteerNet with discriminative selection generates a mask to isolate the target based on its DoA. The generated mask and its complement are used to estimate the target and covariance matrices. A GEV beamformer generates the estimated target signal, and a second one produces the leakage signal. These two signals are fed to PlumberNet, which estimates a fine-grain mask to further improve the target speech quality.}
    \label{fig:pipeline}
\end{figure*}

We consider two speech point sound sources, denoted as $s[n] \in \mathbb{R}$ for the target and $b[n] \in \mathbb{R}$ for the interference, where $n \in \mathbb{N}$ is the sample index.
The expressions $h_m[n] \in \mathbb{R}$ and $h'_m[n] \in \mathbb{R}$ denote the room impulse responses between the sources (target and interference, respectively), where $m \in \{1, 2, \dots, D\}$ is the microphone index and $D \in \mathbb{N}$ is the number of microphones.
The measured signal at each microphone is given by:
\begin{equation}
    x_m[n] = h_m[n] * s[n] + h'_m[n] * b[n],
    \label{eq:x_m}
\end{equation}
where $*$ stands for the linear convolution operator.

A Short-Time Fourier Transform (STFT) with frames of $N \in 2\mathbb{N}$ samples and a hop size of $\Delta N \in \mathbb{N}$ samples generates a time-frequency representation $X_m[l,k] \in \mathbb{C}$ for each microphone signal $x_m[m]$, where $l \in \{0, 1, \dots, L-1\}$ is the frame index (out of $L \in \mathbb{N}$ frames) and $k \in \{0, 1, \dots, N/2\}$ is the frequency bin index.
This can be combined in a single vector $\mathbf{X}[l,k] \in \mathbb{C}^{D \times 1}$:
\begin{equation}
    \mathbf{X}[l,k] = \left[
    \begin{array}{cccc}
    X_1[l,k] & X_2[l,k] & \dots & X_D[l,k]
    \end{array}
    \right]^T,
    \label{eq:Xv}
\end{equation}
where $\{\dots\}^T$ denotes the transpose operator.

As previously demonstrated in \cite{grondin2020gev}, a neural network (e.g. SteerNet) can generate a coarse time-frequency ideal ratio mask $\bar{M}[l,k] \in [0,1]$ from a set of pairs of microphones, based on the direction of arrival of the target sound source, denoted by $\theta$.
\begin{equation}
    \bar{M}[l,k] = f(\theta, \mathbf{X}[l,k])
\end{equation}

\begin{figure}[!ht]
    \centering
    \includegraphics[width=\linewidth]{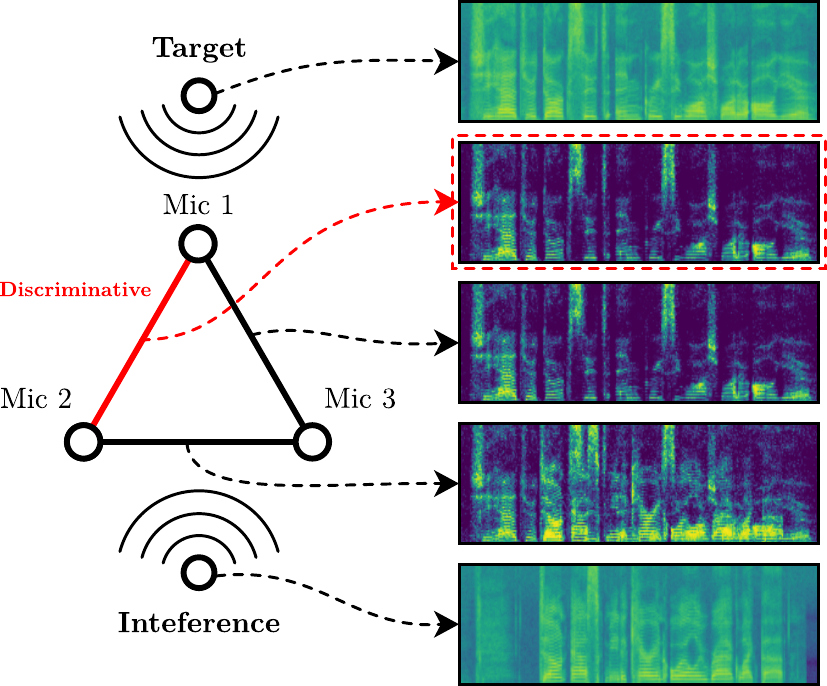}
    \caption{Mask estimation using the most discriminative pair of microphones. SteerNet estimates a mask for each pair, and then the instance with the least amount of time-frequency bin close to a value of $1$ is chosen. In this example, the pair of microphones $2$ and $3$ show an ambiguity as the target and interference have the same time difference of arrival, and the resulting mask shows more time-frequency bins with a value close to $1$ compared to pairs $(1,2)$, and $(1,3)$.}
    \label{fig:masking}
\end{figure}

The estimated mask $\bar{M}[l,k] \in [0,1]$ emphasizes the time-frequency regions dominated by the target speech.
In the former paper \cite{grondin2020gev}, a mask is estimated for each pair of microphones and all masks are averaged to produce $\bar{M}[l,k]$.
However, it is common that some pairs of microphones will perceive both the target and interference sources with the same time difference of arrival (TDoA), and the neural network generates masks that capture both signals.
To overcome this limitation, the mask $\bar{M}[l,k]$ is set to corresponds to the pair of microphones that generate the most discriminative mask, i.e. the mask with the smallest number of active time-frequency bins.
This method is illustrated at figure \ref{fig:masking}.

This mask is used to estimate the SCMs at each frequency bin $k$ for target speech ($\Phi_{\mathbf{SS}}[k] \in \mathbb{C}^{D \times D}$) and interference speech ($\Phi_{\mathbf{BB}}[k] \in \mathbb{C}^{D \times D}$):
\begin{equation}
    \Phi_{\mathbf{SS}}[k] = \sum_{l=1}^{L}{\bar{M}[l,k]\mathbf{X}[l,k]\mathbf{X}[l,k]^H},
\end{equation}
\begin{equation}
    \Phi_{\mathbf{BB}}[k] = \sum_{l=1}^{L}{(1-\bar{M}[l,k])\mathbf{X}[l,k]\mathbf{X}[l,k]^H},
\end{equation}
where $\{\dots\}^H$ stands for the Hermitian operator.

The weights $\mathbf{w}_{target}[k] \in \mathbb{C}^{D \times 1}$ of the Generalized Eigenvalue (GEV) beamformer that estimates the target speech at each $k$, can be calculated using these SCMs by:
\begin{equation}
    \mathbf{w}_{target}[k] = \mathcal{P}\{\Phi^{-1}_{\mathbf{BB}}[k]\Phi_{\mathbf{SS}}[k]\},
    \label{eq:w_target}
\end{equation}
with the principal component operator $\mathcal{P}\{\dots\}$ that extracts the eigenvector associated to the highest eigenvalue.
Note that in case the positive semi-definite matrix $\Phi_{\mathbf{BB}}$ is singular, it can be made positive definite by diagonal loading.

The estimated target speech signal $Y_{target}[l,k] \in \mathbb{C}$ is calculated by:
\begin{equation}
    Y_{target}[l,k] = (\mathbf{w}_{target}[k])^H \mathbf{X}[l,k].
\end{equation}

Although beamforming estimates the target source, there usually remains a part of the undesirable interfering speech (i.e. leakage).
A common strategy consists in predicting another mask from the estimated target signal produced by the beamformer, and in applying this mask as a last postfiltering step.
This approach may however suffer from the source permutation ambiguity, which makes target mask prediction challenging.
In fact, the post-filter neural network can easily confuse both sources if the estimated speech target source has a similar power level than the remaining interfering speech source.
In \cite{yang2023guided}, an arbitrary microphone signal is used as a reference to differentiate the target source from the mixture.
In this work, we demonstrate that using another set of weights to estimate the leakage signal and use it as the reference signal provides better results.
To achieve this, the SCMs are swapped from how they are used in (\ref{eq:w_target}), and the weights that estimate the leakage are similarly computed by:
\begin{equation}
    \mathbf{w}_{leak}[k] = \mathcal{P}\{\Phi^{-1}_{\mathbf{SS}}[k]\Phi_{\mathbf{BB}}[k]\}.
\end{equation}

Here, the same diagonal loading strategy as in (\ref{eq:w_target}) can be applied to $\Phi_{\mathbf{SS}}$ to ensure the matrix is invertible.
The leakage signal $Y_{leak}[l,k] \in \mathbb{C}$ can be then calculated by:
\begin{equation}
    Y_{leak}[l,k] = (\mathbf{w}_{leak}[k])^H \mathbf{X}[l,k].
\end{equation}

A neural network denoted by $g(\dots)$ is used to estimate the postfilter ideal ratio mask $\hat{M}[l,k] \in [0,1]$ using both the estimated target and leakage signals:
\begin{equation}
    \hat{M}[l,k] = g(Y_{target}[l,k], Y_{leak}[l,k]).
\end{equation}

This network estimates the ideal ratio mask for postfiltering, which corresponds to:
\begin{equation}
    M[l,k] = \frac{|Y_{ref}[l,k]|}{|Y_{target}[l,k]|},
\end{equation}
with
\begin{equation}
    Y_{ref}[l,k] = (\mathbf{w}_{target}[k])^H \mathbf{X}_{ref}[l,k],
\end{equation}
where the expression $\mathbf{X}_{ref}[l,k]$ is obtained applying the STFT to (\ref{eq:x_m}), and then using (\ref{eq:Xv}) without the interference signal ($b[n] = 0\ \forall\ n$).

By using both $Y_{target}$ and $Y_{leak}$, the network should be able to solve the possible permutation between target and interference sources. 
For this application, a recurrent neural network (RNN) is fed a tensor made up of $L$ frames and $2(N/2+1)$ frequency bins, which holds the concatenated frames $\log|Y_{target}[l,k]|$ and $\log|Y_{leak}[l,k]|$.
The RNN, implemented using a Gated Recurrent Unit (GRU) with two layers, consists of 256 hidden units at each layer.
The output of the RNN is then fed to a fully connected layer and a sigmoid function.
A dropout layer is also added to the RNN and between the output of the RNN and the fully connected layer, with a dropout probability of $p=0.2$ to reduce overfitting.

The employed loss function is presented in (\ref{eq:lossfunc}), where $\beta=0.25$ ensures proper weighting across all frequencies:
\begin{equation}
\label{eq:lossfunc}
loss = (M[l,k]-\hat{M}[l,k])|Y_{target}[l,k]|^{\beta}.
\end{equation}

Finally, the estimated mask is applied to the magnitude of the GEV beamformer output spectrum:
\begin{equation}
    Z[l,k] = \hat{M}[l,k] Y_{target}[l,k],
\end{equation}
and the inverse STFT is applied to $Z[l,k] \in \mathbb{C}$ to generate the time-domain output $z[n] \in \mathbb{R}$.
Figure \ref{fig:pipeline} shows the complete pipeline previously described, and the full implementation can be accessed online\footnote{https://github.com/balkce/plumbernet}.

\section{RESULTS AND DISCUSSION}

We compare the performance of the proposed method formerly introduced in \cite{yang2023guided}, where the covariance matrices are pre-computed offline during a calibration step. This makes the system less suitable for dynamic environments, whereas here we estimate the covariance matrices based on the direction of arrival of the target.
This makes our approach more convenient to unknown environments as the DoA can be obtained from a video feed or \emph{a priori} knowledge of the target source position.
Moreover, our approach uses the leakage signal as a reference signal, as opposed to an arbitrary microphone signal as in \cite{yang2023guided}.

We created a dataset based on the clean sub-set of the Interspeech 2022 deep noise suppression challenge corpus \cite{reddy2020interspeech}. Rooms of different sizes are simulated with a target speech source and an interference source. 
Room impulse responses (RIRs) are generated using the image method \cite{habets2006room} to simulate sound propagation between the target and interference sources and a pair of microphones spaced randomly between $4$ and $20$ cm, as in \cite{grondin2020gev}.
We then use SteerNet to predict a time-frequency mask to obtain the SCM and the GEV beamformer weights.
In one case, we train a neural network to post-filter the target based on the beamformed signal and a reference microphone.
In the second case, we train the same neural network, but using the beamformed signal and the leakage signal.

A similar dataset is generated for validation, using two microphones randomly spaced in a simulated room with a target source and an interfering source.
A two-microphone GEV beamformer based on the estimated SCM using SteerNet enhances the target source, and the result is fed as the input to the trained neural network for further enhancement.
Table \ref{tab:leak} compares the improvement between the ideal target signal and the estimated signal in terms of Scale-Invariant Signal-to-Distortion Ratio (SI-SDR) \cite{le2019sdr}, Perceptual Evaluation of Speech Quality (PESQ) \cite{rix2001perceptual} and Short-Time Objective Intelligibility (STOI) \cite{taal2011algorithm} scores.
Results clearly demonstrate that using the leakage signal rather than an arbitrary reference microphone improves all metrics.
Note that the goal here is to demonstrate that using a neural network that takes the beamformed signal and the leakage signal (rather than a reference microphone) produces better results with a simple neural network architecture.
This suggests that the information contained in the leakage signal is easier to manipulate than a reference microphone, a mix of target and interference signals.

\begin{table}[]
    \centering
    \renewcommand{\arraystretch}{1.3}
    \caption{SI-SDR, PESQ and STOI with two microphones, with a reference microphone and the proposed PlumberNet that uses the leakage signal as a reference. Results demonstrate the proposed method PlumberNet offers better enhancement for the same neural network architecture.}
    \vspace{10pt}
    \begin{tabular}{c|ccc|ccc}
        \hline
        \hline
        \multirow{2}{*}{Metrics} & \multicolumn{3}{c|}{Reference mic} & \multicolumn{3}{c}{Proposed leakage} \\
         & In & Out & $\Delta$ & In & Out & $\Delta$ \\
        \hline
        SDR & 11.02 & 15.21 & 4.19 & 11.32 & 15.85 & 4.53 \\
        PESQ & 1.765 & 2.611 & .846 & 1.854 & 2.804 & .950 \\
        STOI & 86.0 & 92.2 & 6.2 & 87.0 & 93.5 & 6.5 \\
        \hline
        \hline
    \end{tabular}
    \label{tab:leak}
\end{table}

\begin{figure}[!ht]
    \centering
    \subfloat[ReSpeaker USB\cite{sudharsan2019ai}]{\includegraphics[width=0.33\linewidth]{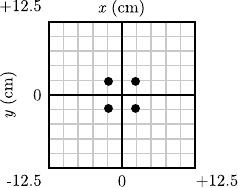}}
    \subfloat[ReSpeaker Core]{\includegraphics[width=0.33\linewidth]{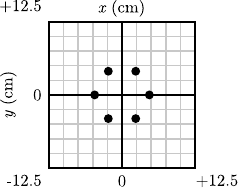}}
    \subfloat[Matrix Creator \cite{haider2019system}]{\includegraphics[width=0.33\linewidth]{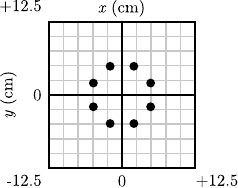}}\\\vspace{-5pt}
    \subfloat[Matrix Voice]{\includegraphics[width=0.33\linewidth]{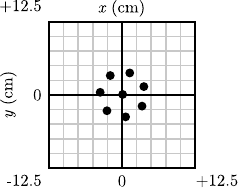}}
    \subfloat[MiniDSP UMA \cite{agarwal2018opportunistic}]{\includegraphics[width=0.33\linewidth]{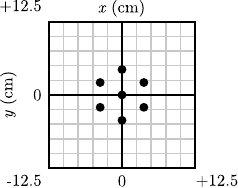}}
    \subfloat[MS Kinect \cite{pei2013sound}]{\includegraphics[width=0.33\linewidth]{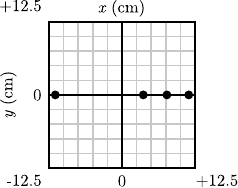}}
    \vspace{-5pt}
    \caption{Microphone array geometries used for testing \cite{grondin2020gev}.}
    \label{fig:micarray_geometry}
\end{figure}

To evaluate with unseen array geometries, another validation set is created with RIRs estimated for randomly selected commercially-available geometries (shown in Fig. \ref{fig:micarray_geometry}).

Since the SCM is usually unknown in practical scenarios, it should be estimated online from the target DoA.
The proposed SteerNet in \cite{grondin2020gev} estimates a mask for each pair of microphones, and these are averaged to produce a single mask used to compute the SCM.
Here, we also explore the use of the most discriminative mask as illustrated in Fig. \ref{fig:masking}.
Table \ref{tab:discriminative} shows that using a discriminative mask improves performances.
This suggests that averaging all pairwise masks introduces leakage during SCM estimation, whereas the discriminative mask leads to a SCM closer to the oracle value.

It is expected that the results shown in Table \ref{tab:discriminative} are lower than those shown in Table \ref{tab:leak}, since unseen array geometries are used, instead of the trained pair geometry.

\begin{table}[]
    \centering
    \renewcommand{\arraystretch}{1.3}
    \caption{SI-SDR, PESQ and STOI with arbitrary microphone array shapes, with the proposed PlumberNet, using masks obtained from averaging or using the most discriminative pair of microphones. Results demonstrate using the most discriminative pair of microphones improves the overall enhancement performances.}
    \vspace{10pt}
    \begin{tabular}{c|ccc|ccc}
        \hline
        \hline
        \multirow{2}{*}{Metrics} & \multicolumn{3}{c|}{Averaging pairs} & \multicolumn{3}{c}{Discriminative} \\
         & In & Out & $\Delta$ & In & Out & $\Delta$ \\
        \hline
        SDR & 11.23 & 12.74 & 1.51 & 13.07 & 14.93 & 1.86 \\
        PESQ & 2.046 & 2.521 & .475 & 2.203 & 2.772 & .569 \\
        STOI & 87.0 & 89.0 & 2.0 & 89.9 & 92.4 & 2.5 \\
        \hline
        \hline
    \end{tabular}
    \label{tab:discriminative}
\end{table}

The results shown here are for a simple RNN architecture, rather than a more fine-tuned and dedicated model such as GSENet \cite{yang2023guided}.
However, it is worth mentioning that the average SI-SDR reported in \cite{yang2023guided} at the beamforming stage (with a speech interference) was 7.5 dB, and at the final stage it was 11.8 dB. Compared to the SDR obtained using the most discriminative mask, shown in Table \ref{tab:discriminative}, it can be observed that GSENet was outperformed by the proposed approach, which uses a smaller network, and without the need of oracle SCMs.

\section{CONCLUSION}
\label{sec:conclusion}

Deep-learning-based speech enhancement has recently grown in interest, as beamforming-based techniques tend to leak interferences in the resulting audio output.
Thus, hybrid solutions have now been applied that use a spatial filter as the input of neural network that serves as a post-filter.
Some of these solutions also employ a second channel to inform the neural network of what is the source of interest and what is the leaked interference.
A recent work uses an arbitrarily chosen microphone input as the second channel, while still being beholden to \textit{a priori} knowledge of the spatial filter coefficients.

In this work, a leakage signal is used as the second channel, which is calculated using a swapped version of spatial correlation matrices which, in turn, are estimated from the microphones using the SteerNet approach.
Results show that the performance increase between the spatial filter output and the trained model is greater when using the leakage signal than when using an arbitrarily chosen microphone.
Additionally, results also showed that using the spatial correlation matrix from the most discriminative microphone pair provides a greater increase in performance than when using the average spatial correlation matrix from all microphone pairs.

For future work, other types of spatial filters are to be explored, as well as other calculations of the final spatial correlation matrix, such as averaging the most discriminant pairs.

\clearpage

\bibliographystyle{IEEEbib}
\bibliography{refs}

\end{document}